\documentstyle[twocolumn,prl,aps,epsfig,floats]{revtex}
\begin{document}
\draft
\twocolumn[\hsize\textwidth\columnwidth\hsize\csname @twocolumnfalse\endcsname

\title{
$SU(4)$ Fermi Liquid State and Spin Filtering in a Double Quantum Dot System}
\author{
L\'aszl\'o Borda$^{1,3}$, Gergely Zar{\'a}nd$^{2,3}$, 
Walter Hofstetter$^2$, B.I. Halperin$^2$, and Jan von Delft$^1$
}

\address{
$^1$Sektion Physik and Center for Nanoscience, 
LMU M\"unchen, 80333 M\"unchen, Theresienstr. 37.\\
$^2$Lyman Physics Laboratory, Harvard University, Cambridge, MA \\
$^3$Research Group of the Hungarian Academy of Sciences, Institute of Physics,
TU Budapest, H-1521 
}
\date{\today}
\maketitle

\begin{abstract}
We  study a symmetrical double quantum dot (DD) system with strong 
capacitive inter-dot coupling using renormalization group methods.
The  dots are  attached to separate  leads,  and there can be a 
weak  tunneling between them.  In the regime where there is a single 
electron on the DD the low-energy behavior is
characterized by an $SU(4)$-symmetric Fermi liquid theory with entangled
spin and charge Kondo correlations and a phase shift $\pi/4$. 
Application of an external magnetic field gives rise to a large  
magneto-conductance
and a crossover to a purely charge Kondo state 
in the charge sector with  $SU(2)$ symmetry.  
In a four lead setup we find perfectly spin polarized transmission.
\end{abstract}
]
\narrowtext

{\em Introduction.---}
Quantum dots are one of the most basic building blocks of 
mesoscopic circuits\cite{Kouwenhoven}. In many respects quantum dots 
act  as  large complex atoms coupled to conducting 
leads that are used to study transport.  
The physical properties of these dots depend essentially 
on the level spacing and  precise form of the coupling
to the leads: They can exhibit  Coulomb 
blockade phenomena \cite{Wilkins}, 
build up  correlated Kondo-like states of various 
kinds\cite{Goldhaber-Gordon,vanderWiel,Matveev}, 
or develop conductance fluctuations. 

\begin{figure}
\begin{center}
\epsfxsize=5.0cm
\epsfbox{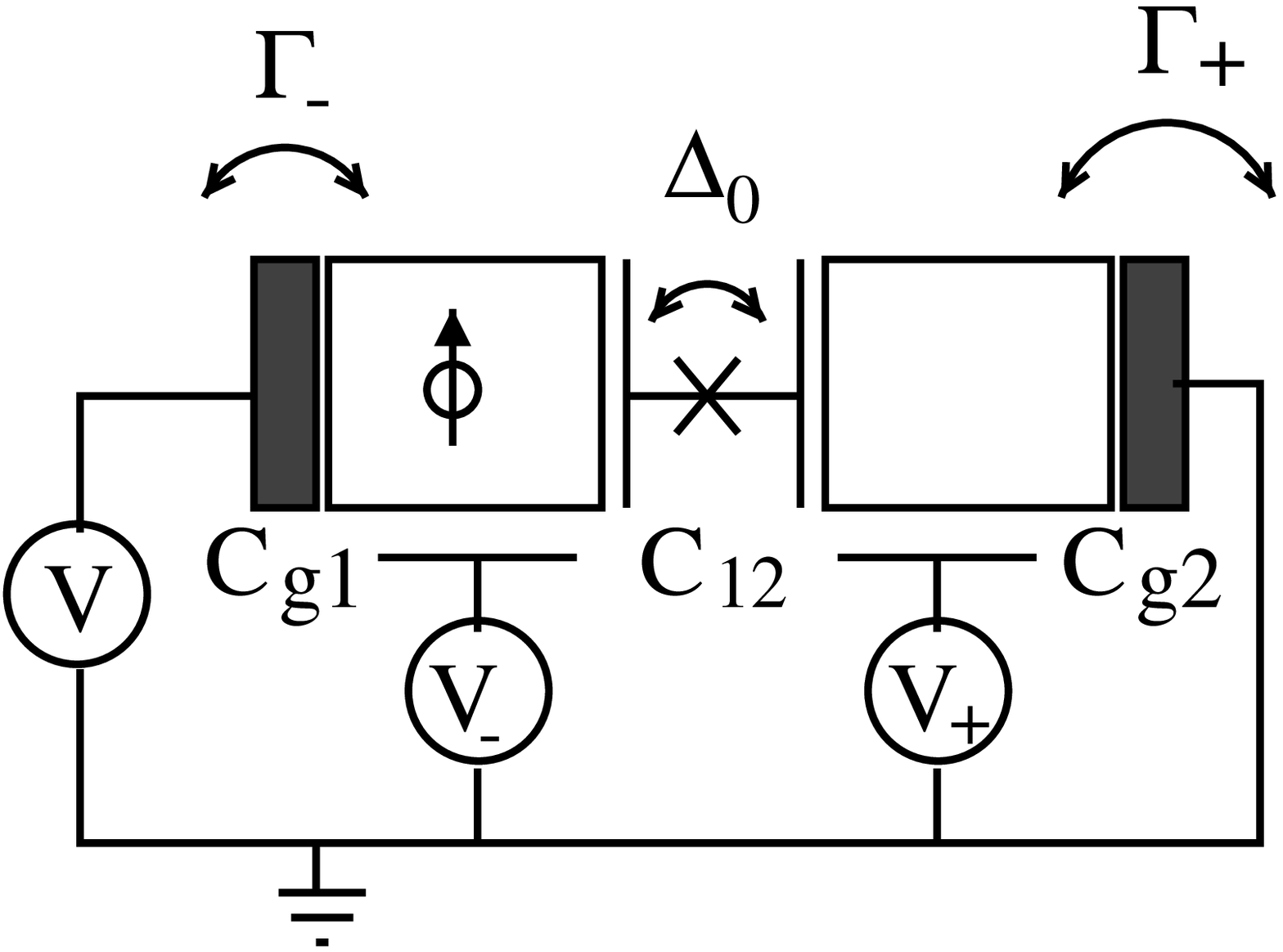}
\vskip0.4truecm
\epsfxsize=4.0cm
\epsfbox{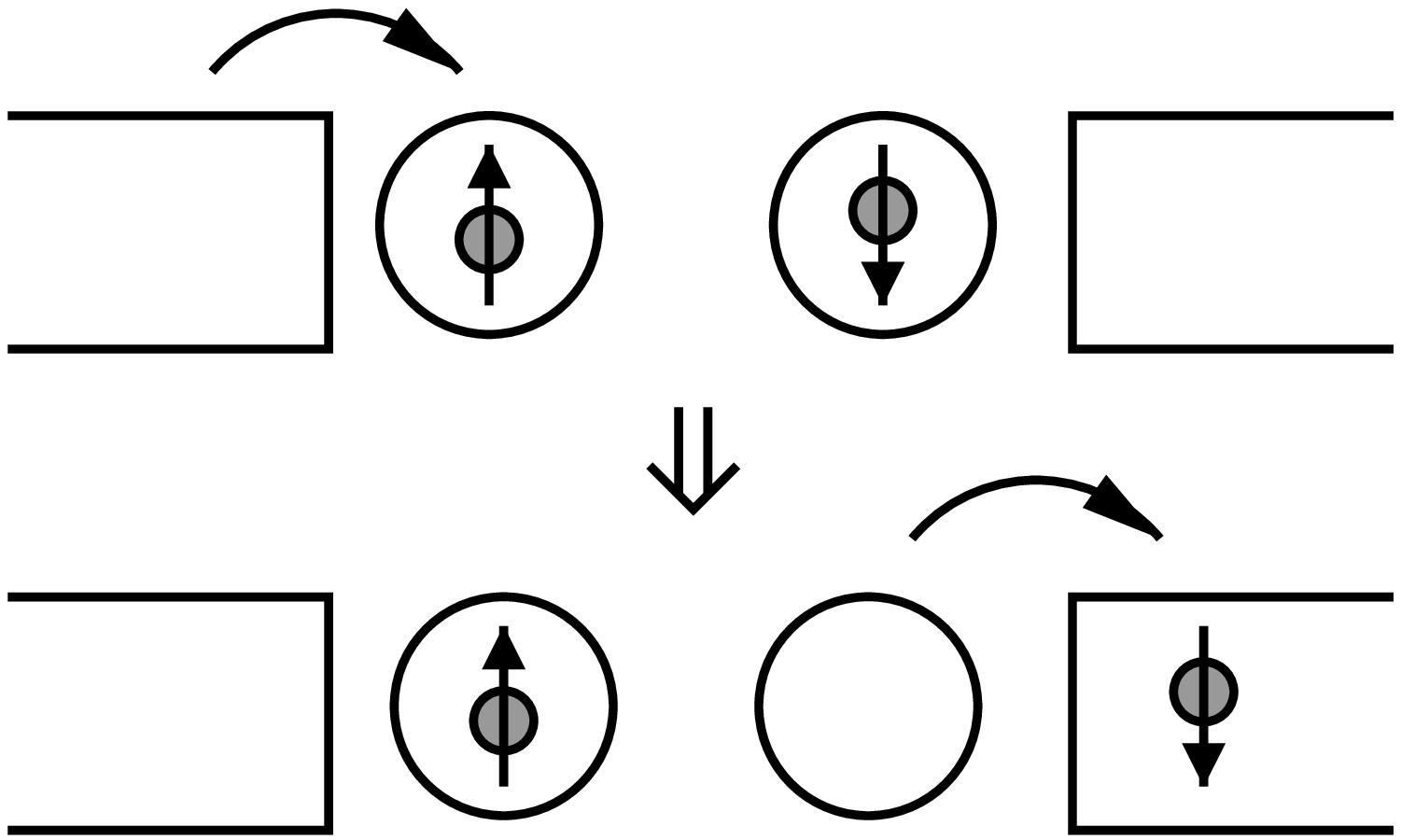}
\end{center}
\caption{\label{fig:2dot}
Top: Schematics of the DD device. Bottom: Virtual process
leading to 'spin-flip assisted tunneling' as described in 
Eq.~(\protect{\ref{eq:assist}})
}
\end{figure}

The simplest mesoscopic circuits that go beyond  
single dot devices in their complexity are  double dot  (DD) devices
(see Fig.~\ref{fig:2dot}). These 'artificial molecules'
have been extensively studied both
theoretically~\cite{Ruzin,Golden,Matveev_2,Izumida,Pohjola,Aguado}
and experimentally~\cite{Waugh,Molenkamp,Blick,Oosterkamp}: 
They may give rise to   stochastic Coulomb  blockade\cite{Ruzin} and 
peak splitting\cite{Golden,Waugh},
can be used as  single electron pumps\cite{Kouwenhoven}, 
 were proposed to measure  high frequency quantum noise\cite{Aguado}, 
and are   building blocks for more complicated 
mesoscopic devices such as  turn-stiles or cellular automata \cite{Toth}. 
DD's also have  interesting 
degeneracy points where quantum fluctuations may 
lead to unusual strongly correlated states\cite{Boese}.

In the present paper we focus  
our attention to small semiconducting DDs with 
large inter-dot capacitance \cite{Pohjola,Boese}. 
We consider the  regime where 
the gate voltages  $V_{\pm}$ are  such that 
the lowest lying charging states,   $(n_+,n_-) = (0,1)$ and  $(1,0)$, are 
almost degenerate: $E(1,0)-E(0,1)\approx  0$
[$n_\pm = $ $\#$ of extra electrons on dot '$\pm$',
and $E(n_+,n_-)$ is measured from the common chemical potential of the 
two leads]. We consider the simplest, most common case 
where the states $(1,0)$ and  $(0,1)$ have both spin 
$S=1/2$, associated with the extra electron on the dots. 
Then at energies below the charging energy of the DD, 
${\tilde E}_C\equiv  \min\{ E(1,1)-E(0,1), E(0,0)-E(0,1) \}$, 
the dynamics of the DD is restricted to the subspace 
$\{S^z = \pm 1/2; \; n_+-n_- = \pm 1\}$.

As we discuss below, quantum fluctuations between these four quantum states 
of the DD generate an unusual strongly correlated Fermi liquid state, 
where  the  spin and charge degrees of freedom of the DD are totally 
entangled. We show that this state possesses 
an $SU(4)$ symmetry 
corresponding 
to the total internal degrees of freedom of the DD, and is 
characterized by a phase shift $\delta = \pi/4$. 
This phase shift can be measured 
by integrating the DD device in an Aharonov-Bohm interferometer\cite{Yacoby}.
Application of an external  field on the DD suppresses spin fluctuations. 
However, charge fluctuations are 
unaffected by the magnetic field
and still give rise to a  Kondo  effect in the  charge (orbital) 
sector \cite{Pohjola,Boese,Weis}.
{  We show that in a four lead setup this latter state gives 
rise to an almost {\em totally spin polarized}  current through the DD
with a field-independent conductance $G\approx e^2/h$. 
The conductance {\em across} the dots, on the other hand, 
shows a large {\em negative magneto-resistance} at $T=0$ temperature.}

{\em Model.---} 
Let us first discuss the arrangement in Fig.~\ref{fig:2dot}.
At energies below ${\tilde E}_{C}$ we describe   
the isolated DD in terms of the {\em orbital pseudospin}
 $T^z \equiv {(n_+ - n_-)/  2} = \pm \frac12$:
\begin{equation}
H_{\rm dot} = -\delta E  \; \; T^z -  t \; T^x - B\;S^z\;.
\end{equation}
The term proportional to $T^z$ describes 
the {\em energy difference} of the two charge states [$\delta E
\equiv  E(1,0)-E(0,1) \sim V_+ - V_-$ for a fully symmetrical  system],  
while  $t \ll {\tilde E}_{C}$  is the tunneling amplitude  between them.
The last term stands for the Zeeman-splitting due to an applied local 
magnetic field in the $z$ direction.
We are interested in the regime, where --- despite the large capacitive
coupling,--- the tunneling between the dots is small. 
Furthermore, one needs a large enough single particle 
level spacing $\Delta$ on the dots. 
Both conditions can be satisfied by making small 
dots \cite{Davidprivate}, which are close together 
or capacitively coupled to a common top-gate electrode \cite{Westervelt}.

The leads are described by the Hamiltonian:
\begin{equation}
H_{\rm leads} = \sum_{|\varepsilon| < D} \epsilon \;a^\dagger_{\varepsilon \sigma +}
a_{\varepsilon \sigma +} + \sum_{|\varepsilon| < D}
\epsilon \;a^\dagger_{\varepsilon \sigma -}
a_{\varepsilon \sigma -}\;,
\end{equation}
where  $a^\dagger_{\varepsilon \sigma +}$ 
($a^\dagger_{\varepsilon \sigma -}$) creates an electron in the right 
(left) lead with energy $\varepsilon$ and spin $\sigma$, 
$D\sim {\rm min}\{ {\tilde E}_{C}, \Delta\} \equiv 1$ is a cut-off, 
and  $\{a^\dagger_{\varepsilon \sigma \tau},
a_{\varepsilon' \sigma' \tau'}\}= \delta_{\tau\tau'}
\delta_{\sigma\sigma'} \delta(\varepsilon-\varepsilon')$.

To determine the effective DD -- lead coupling we 
have to consider virtual charge fluctuations 
to the excited states with $n_+ + n_- = 0$ and 2,
generated by tunneling from the leads to the dots.
By second order perturbation theory  
in the lead-dot tunneling  we obtain the following   effective Hamiltonian:
\small
\begin{eqnarray}
&&H_{\rm Kondo} = 
{1\over2}J_+ P_+  {\vec S}  (\psi^\dagger
{\vec \sigma }\;p_+ \psi )  + 
{1\over2}
J_- P_-  {\vec S}  (\psi^\dagger
{\vec \sigma }\; p_- \psi ),
\label{eq:Kondo}
\\
&&H_{\rm assist} = 
Q_{\perp}\left[ 
T^+  {\vec S}  (\psi^\dagger
{\vec \sigma } \tau^- \psi ) \;\; + \;\; \mbox{h.c.}\right], 
\label{eq:assist}
\\
&&H_{\rm orb}= {1\over2}\Big(V_zT^z(\psi ^\dagger \tau^z \psi)
+V_\perp[T^+(\psi ^\dagger \tau^- \psi) + \mbox{h.c.}]\Big).
\label{eq:orbital}
\end{eqnarray}
\normalsize
where  $ \psi_{\sigma\tau} = \int d\varepsilon 
\;a_{\varepsilon\sigma\tau}$,  and ${\vec \sigma}$   
 and ${\vec \tau}$  denote the spin and  {\em orbital pseudospin} 
of the electrons ($\sigma = \uparrow,\downarrow$; $\tau = \tau^z=\pm1$).
The operators  $P_\pm = (1\pm2 T^z)/2$ and $p_\pm = (1\pm \tau^z)/2$
project out the   DD states $(1,0)$ and $(0,1)$, and the right/left 
lead channels, respectively.

In the limit of small dot-lead tunneling the dimensionless 
exchange couplings are
$J_{\pm} \sim \Gamma_\pm/{\tilde E}_C$ with  $\Gamma_\pm$ the 
tunneling rate to the right (left) lead \cite{footnote2}.
The 'spin-flip assisted tunneling'  $Q_{\perp}  \sim \sqrt{\Gamma_+ 
\Gamma_-}/{\tilde E}_C$ in  Eq.~(\ref{eq:assist}) gives simultaneous  
spin- and  pseudospin-flip scattering and is produced by virtual 
processes depicted in the lower part of Fig.~\ref{fig:2dot}, while the
spin-independent parts of such virtual processes lead to the orbital
Kondo term in Eq.~(\ref{eq:orbital}) with similar amplitudes.

We first focus on  
the case of a fully symmetrical  DD. 
Then the sum of Eqs.~(\ref{eq:Kondo}) and (\ref{eq:assist}) 
can be rewritten as:
\begin{eqnarray}
&& H_{\rm Kondo}+ H_{\rm assist} = {1\over2}{J}  
 {\vec S}  (\psi^\dagger  {\vec \sigma } \psi)\; 
\label{eq:rewritten} \\
&+& 
{Q_{z}}  T^z  {\vec S}  
(\psi ^\dagger \tau^z {\vec \sigma } \psi) 
+ 
{Q_{\perp}}\left( 
T^+  {\vec S}  (\psi^\dagger \tau^-
{\vec \sigma } \psi ) + \mbox{h.c.}\right)\;,\nonumber
\end{eqnarray}
where $J= Q_z=(J_+ +J_-)/4$. The couplings in 
Eqs.~(\ref{eq:Kondo}-\ref{eq:orbital}) are  not entirely 
independent, but are related by the constraints 
$V_\perp = Q_\perp$ and $J = Q_z$.

{\em Scaling Analysis.---}
The perturbative scaling analysis  follows that of a related model 
in  Ref.~\cite{Zarand}.
In the perturbative RG one performs the scaling 
by integrating out conduction 
electrons with energy larger than a scale scale $\tilde D \ll D $, 
and thus obtains an effective Hamiltonian 
that describes the physics at energies $\tilde D$. 
For 
$\delta E=t=B=0$,
in the leading logarithmic 
approximation we find that  all couplings diverge at the Kondo temperature 
$T_K^{(0)}$, where the perturbative scaling breaks down.
Nevertheless, the {\em structure} of the divergent couplings 
suggests that  at low energies  $J=V_\perp = V_z = Q_\perp = Q_z$. 
Thus at small energies, --- apart from a trivial potential scattering --- 
the effective model is a remarkably 
simple  SU(4) symmetrical  exchange model:
\begin{equation}
H_{\rm eff}(T\to0) = \tilde J \sum_{\alpha,\beta = 1,..,4} \psi_\alpha^\dagger 
\psi_\beta \; |\beta\rangle \langle \alpha|\;,
\label{eq:H_eff}
\end{equation}
where the index $\alpha$ labels the four combinations of 
possible spin and pseudospin  indices, and the $|\alpha\rangle$'s denote the 
four states of the DD. This indeed can be more rigorously proven
using strong coupling expansion, conformal field theory, 
and large $f$ (flavor)  
expansion techniques \cite{Nozieres_Blandin,Ye,M_state_PRL}, and is also 
confirmed by our numerical computations.

{\em Numerical Renormalization Group (NRG).---}
To access the low-energy physics of the 
DD, we  used Wilson's NRG approach\cite{NRG}.
In this method one defines a series of 
rescaled Hamiltonians, $H_N$, related by the relation \cite{NRG}:
\begin{eqnarray}
{H}_{N +1} & \equiv &\Lambda^{1/2}{H}_N+\sum_{\sigma\tau}
\xi_{N}\left(f^{\dagger}_{N,\sigma\tau}f_{N+1,\sigma\tau}+ \mbox{h.c}
\right), 
\label{eq:rec}
\end{eqnarray}
where $f_{0\sigma\tau} = \psi_{\sigma\tau}/\sqrt{2}$ and 
$H_0 \equiv 2\Lambda^{1/2}/(1 + \Lambda)\; H_{\rm int}$ 
with $\Lambda\sim3$ as discretization parameter, and
$\xi_N\approx 1$. {  (For the definition of $f_N$
see Ref.~\cite{NRG}.)}  We have defined 
$H_{\rm int} = H_{\rm dot} + H_{\rm Kondo} + 
H_{\rm assist} + H_{\rm orb}$.
The original Hamiltonian is related to 
the $H_N$'s as
$
H = \lim_{N\to\infty} \omega_N H_N
$ 
with  $\omega_N = \Lambda^{-(N+1)/2}(1 + \Lambda)/2$.
Solving Eq.~(\ref{eq:rec}) iteratively 
we can then use the eigenstates of  $H_N$ to calculate physical 
quantities at a scale $T,\omega \sim \omega_N$. 

{\em Results.---} First let us consider the case $H_{\rm dot} = 0$.

{\em Fixed point structure.---}
The finite size spectrum produced by the  NRG procedure  
contains a lot of information. 
Among others, we can  identify the structure of the low-energy 
effective Hamiltonian  from it \cite{NRG},  and also 
determine all  scattering phase shifts.

In particular, we find that for $\delta E = t = B = 0$ the 
entire finite size  spectrum can be understood as a sum 
of four independent, spinless chiral fermion spectra
with phase shifts  $\delta=\pi/4$.  This phase shift is characteristic for 
the $SU(4)$ Hamiltonian, Eq.~(\ref{eq:H_eff}),  and simply follows from  
the Friedel sum rule \cite{Nozieres_Blandin}. Application of an external magnetic 
field $B$ to the DD  gradually shifts $\delta$ to the values
$\delta_\uparrow \to  \pi/2$ and $\delta_\downarrow  \to 0$
\cite{future}.

{\em Spectral functions.---}
To learn more about the dynamics of the DD we computed 
at $T=0$ 
the 
 spin spectral function $\varrho_S^z(\omega) = 
-(1/ \pi){\rm Im} \{\chi_{S}^z(\omega)\}$, 
and pseudospin spectral 
function
 $\varrho_\tau^y(\omega) = 
-(1/ \pi){\rm Im}\{\chi_{\tau}^y(\omega)\}$
by the  density matrix NRG method~\cite{Hofstetter}.

\begin{figure}
\epsfxsize=6cm
\begin{center}
\epsfbox{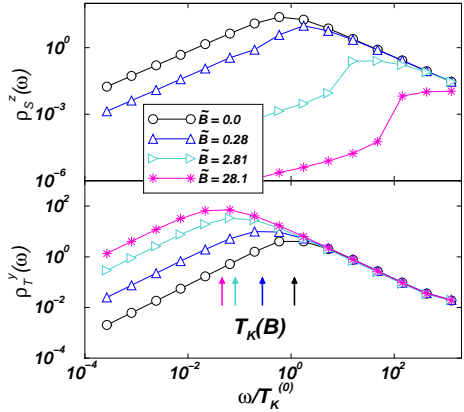}
\end{center}
\caption{
\label{fig:sp_func}
$T=0$ spin and pseudospin spectral  functions for 
$J=Q_z=V_z= 0.14$,  $V_{\perp}=Q_{\perp}=0.13$, 
and various values of $\tilde B \equiv B/T_K^{(0)}$. 
For $B=0$ both the
spin and pseudospin spectral functions 
exhibit $\sim\omega$ behavior below the Kondo temperature
$T^{(0)}_K\sim 10^{-3}$. Applying a magnetic field the situation
changes: The $B>T^{(0)}_K$ magnetic field destroys the spin Kondo correlations 
and leads to a purely orbital Kondo-effect.
}
\end{figure}

At $B=0$ the various spectral functions exhibit a peak at 
the same energy, $T^{(0)}_K$, corresponding to the formation of the $SU(4)$
symmetric state (see Fig.~\ref{fig:sp_func}). Below $T^{(0)}_K$ 
all spectral functions become linear,  characteristic to a Fermi 
liquid state with local spin and pseudospin susceptibilities
$\chi_S \sim \chi_T \sim 1/T^{(0)}_K$,
where the $SU(4)$ ''hyper-spin'' of the dot electron (formed by 
$\{\uparrow+,\downarrow+,\uparrow-,\downarrow-\}$ components) is
screened by the lead electrons.

Now let us consider the case $H_{\rm dot} \ne 0$.
In a large magnetic field,  
$T^{(0)}_K\ll B$, spin-flip processes are suppressed: The spin
spectral function therefore shows only a Schottky anomaly at $\omega\sim B$.
Nevertheless, the couplings $V_\perp$ and $V_z$ still generate a purely 
orbital Kondo state  
in the spin channel with the same orientation as  the  DD spin,
with  a reduced Kondo temperature $T_K(B)< T^{(0)}_K$, and a corresponding 
phase shift $\delta_\uparrow = \pi/2$.

Due to the  spin-pseudospin symmetric structure
of the Hamiltonian, Eq.~(\ref{eq:rewritten}), the opposite
effect occurs for a large $\delta E$: In that limit the 
charge is localized on one side of the DD, charge fluctuations are 
suppressed,  and the system scales to a spin Kondo problem. 
A large tunneling, $t > T_K^{(0)}$ is also expected to lead 
to a somewhat similar effect, though the conductance through the 
DD behaves very differently in the two cases \cite{future}.
 
\begin{figure}
\epsfxsize=7.0cm
\begin{center}
\epsfbox{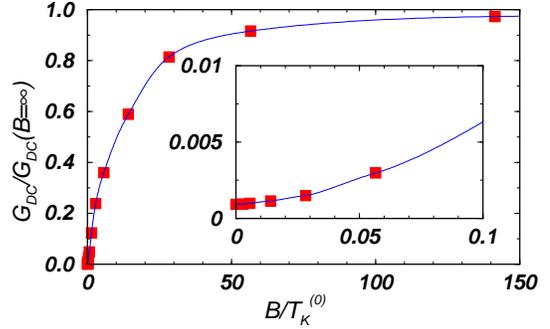}
\end{center}
\caption{
\label{fig:sigma}
The $T=0$ conductance of DD system at $\omega=0$
for $J=Q_z= 0.14$,  $V_{\perp}=Q_{\perp}=0.13$, 
$V_z=0.14$ and different magnetic field values. 
Inset: small $B$ limit of the conductance.
}
\end{figure}

{\em DC Conductivity.---} First we focus on the 
conductivity {\em across the DD} assuming 
a small tunneling $t$.
Then we can assume that the two dots are in equilibrium with the 
leads connected to them, and we can compute the induced current
perturbatively in $t$. 
A simple calculation yields the following formula:\cite{comment} 
\begin{equation}
G = {{2\pi^2 e^2}\over h}
\;\;  {t^2}\;\; \lim_{\omega\to 0} {\varrho_\tau^y(\omega)
\over \omega}
\;.
\label{sigma}
\end{equation}
The normalized DC conductance at $T=0$ temperature 
is shown in  Fig~\ref{fig:sigma}.  
Below the orbital Kondo temperature 
$\varrho_\tau^y(\omega)\sim \omega/T^2_K(B)$, 
leading to a dimensionless conductance  $ \sim [t/T_K(B)]^2$. 
However, $T_K(B)$ strongly decreases with 
increasing $B$ implying a {\em large negative 
magneto-resistance} in the $T=0$ DC conductance. This effect 
is related to the correlation between spin 
and orbital degrees of freedom. 
We have to emphasize that the  simple considerations above only apply 
in the regime $t\ll T_K(B)$. For larger values of $t$
a more complete calculation is needed.

\begin{figure}
\begin{center}
\epsfxsize=7.0cm
\epsfbox{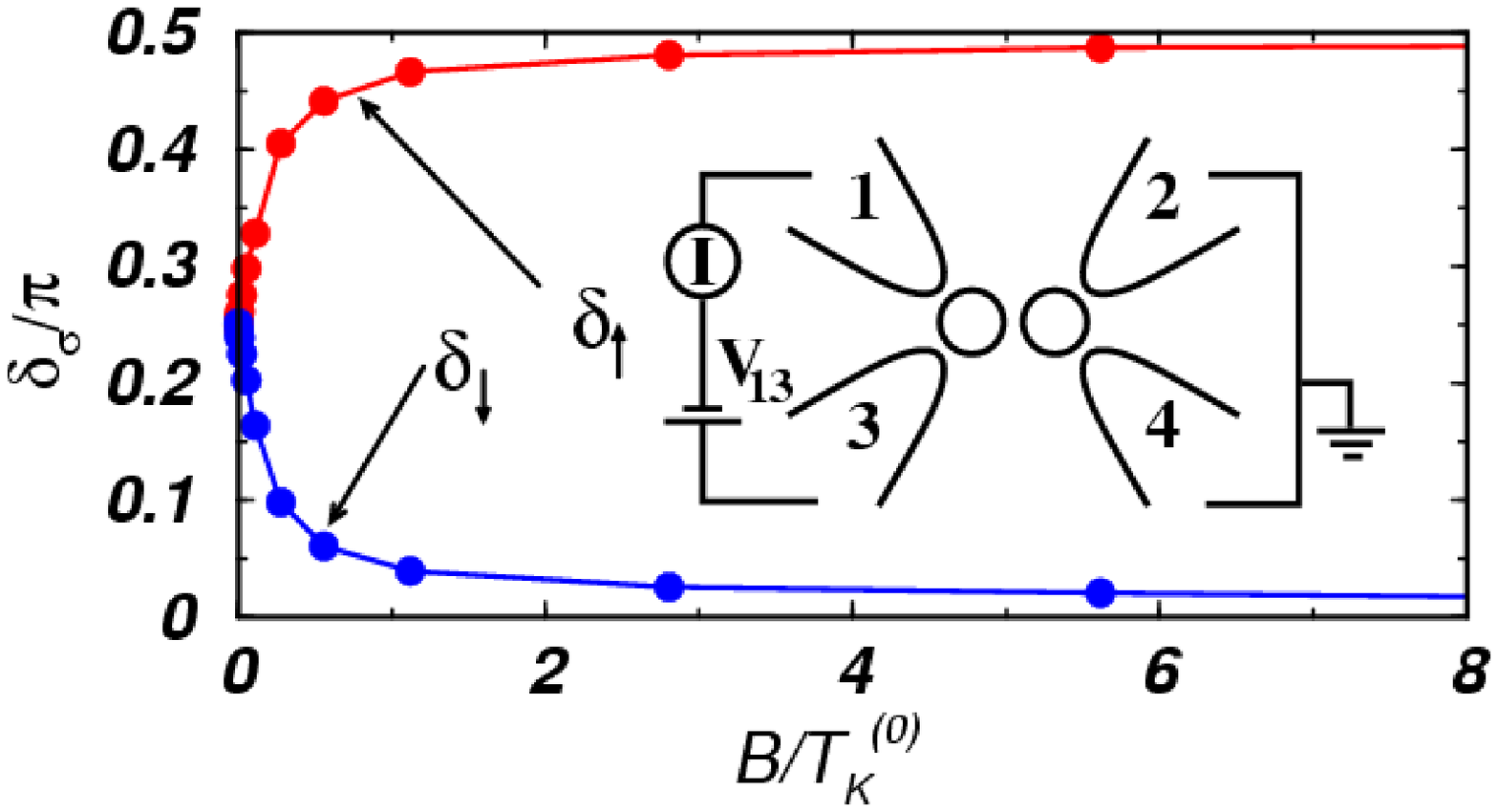}
\epsfxsize=7.0cm
\epsfbox{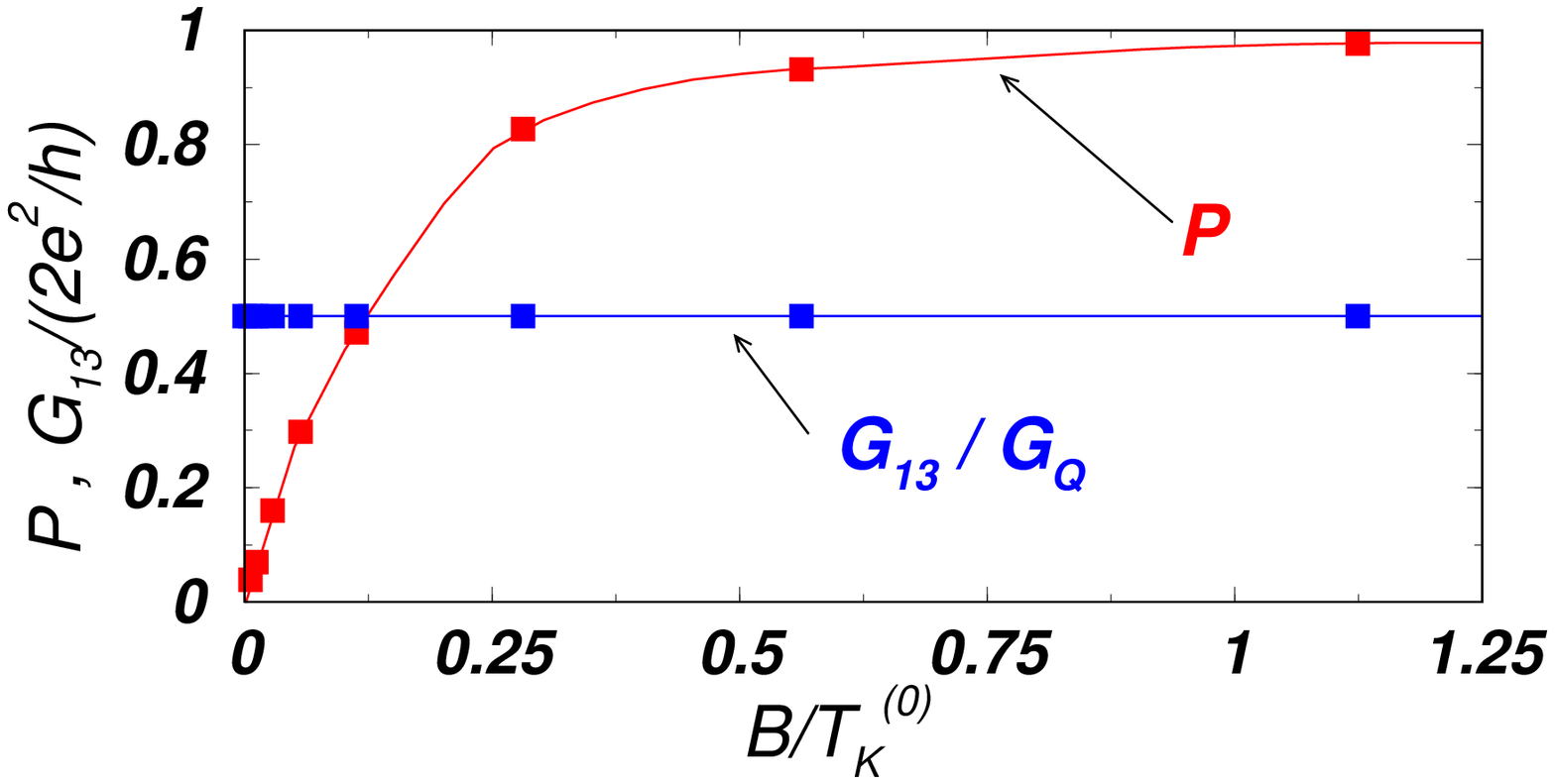}
\end{center}
\caption{
\label{fig:polarization} Top: Magnetic field dependence of the 
phase shifts for $t=\delta E \approx 0$.  Bottom: Corresponding 
$T=0$ dimensionless conductance and spin polarization of 
the current in the four lead setup shown in the Top inset.
} 
\end{figure}

Having extracted the phase shifts from the NRG spectra,
we can construct the scattering matrix in more general geometries too 
and compute the $T=0$ conductance using the Landauer-Buttiker  
formula \cite{future,Pustilnik,Buttiker}. 
In the perfectly symmetrical two terminal four lead 
setup of Fig.~\ref{fig:polarization} with $\delta E =t=0$, e.g.,  
the DC conductance is 
$G_{13} =  {1\over 2} G_Q (\sin^2(\delta_\downarrow(B)) 
+ \sin^2(\delta_\uparrow(B)))$,  where $G_Q = 2e^2/h$ is the 
quantum conductance. By the Friedel sum rule $\delta_\uparrow(B) 
= \pi/2 - \delta_\downarrow(B)$, and thus $G_{13}(T=0) =  G_Q/2$, 
independently of $B$. However, the polarization of the transmitted current, 
$P=2 \sin^2(\delta_\uparrow) - 1$  tends rapidly 
to  1 as $B > T_K^{(0)}$, and the DD thereby acts as a {\em perfect 
spin filter} at $T=0$ with $B>T_K^{(0)}$, and could also serve
as a spin pump. For a typical $T_K\approx 0.5K$ and a $g$-factor
$g\approx -0.4$ as in GaAs, {\it e.g.}  a field of $2.5 T$ 
would give a $97 \%$ polarized current. 
This efficiency is comparable to other spin filter 
designs \cite{Potok}. By lowering $T_K$, much higher 
polarizations could be obtained.

{\em Robustness.---} 
Since the spin $S^\alpha$ and pseudospin $T^\alpha$  
are both marginal operators at the $SU(4)$ fixed 
point \cite{Ye}, 
we conclude that the $SU(4)$ behavior is stable 
{  in the sense that a small but finite value of
$\delta E$, $B$, $t\ll T_K^{(0)}$ will lead only to small 
changes in physical properties like the phase shifts.}
The anisotropy of the  couplings 
is also irrelevant in the RG sense \cite{Ye,M_state_PRL}, 
and the role of $J_-\ne J_+$ symmetry breaking is 
only to renormalize the bare value of $\delta E$, which is a marginal 
perturbation itself. Therefore the $SU(4)$ Fermi liquid
state is {\em robust} under the conditions discussed in 
the Introduction.

{\em Experimental accessibility.---} 
For our scenario  it is crucial to have large enough 
charging energy and level spacing  
${\tilde E}_{C}, \Delta >T_K^{(0)} > t$.
With today's technology it is possible to reach 
$\Delta\sim 2-3{\rm K}$.  
The dot-dot capacitance  (and thus ${\tilde E}_{C}$ \cite{Golden}) 
can  be  increased 
by changing the shape of the gate electrode
separating the dots, using a columnar geometry as in 
Refs.~\cite{Weis,Tarucha}, where the
two-dimensional  dots are  placed on the top of  
each other, or  placing an additional electrode on the top of the 
DD device\cite{Westervelt}. 
We could not find a closed expression for $T_K^{(0)}$ 
in the general case. However, for a symmetrical
DD $J\approx V_\perp \approx V_z 
\approx Q_\perp \approx Q_z \sim \Gamma /2\pi {\tilde E}_C$,
provided that fluctuations to the $(0,0)$ state give the dominant 
contribution. Then we obtain  $T_K^{(0)} \approx 
D e^{-1/4J}$ and $T_K(B=\infty) \approx {\rm cst.}\; {[T_K^{(0)}]}^2/D$.
Thus the value of $J$ and thus  $T_K^{(0)}$ can be tuned 
experimentally to a value similar to the single dot experiments. 
Indeed, an orbital Kondo effect has recently been observed~\cite{Weis}.

{\em Summary.---}
We have studied a DD system with large 
capacitive coupling close to its degeneracy
point, in the Kondo regime. 
Using both scaling arguments and a non--perturbative NRG analysis, 
we showed that the simultaneous appearance of  
the Kondo effect in the spin and charge sectors  
results in an $SU(4)$ Fermi liquid ground state with a phase shift
$\pi/4$. Upon applying an external magnetic field,  
the system crosses over to a purely charge Kondo state 
with a lower $T_K$.
In a four--terminal setup, the DD could thus be used as a spin filter 
with high transmittance. 
We further predict a large serial magneto-conductance 
at $T=0$. The $SU(4)$ behavior in this system is robust, 
and is experimentally accessible.

{\em Acknowledgements}:
We are grateful to T. Costi, K. Damle, and D. Goldhaber-Gordon 
for discussions. This research has been supported  by
NSF Grants Nos. 
DMR-9981283 
and Hungarian Grants No. 
OTKA F030041,  
T038162, 
and  N31769. 
W.H. acknowledges financial support from the 
German Science Foundation (DFG).


\begin{thebibliography}{99}
\bibitem{comment} This formula can also be used at $T>0$.
\end{thebibliography}
\end{document}